\documentclass{PoS}
\usepackage{amsmath}

\title{Perturbative renormalization of $\Delta F = 2$ four-fermion operators with the chirally rotated 
Schr\"odinger functional }

\ShortTitle{Perturbative renormalization of four-fermion operators in the $\chi$SF}

\author{Mattia Dalla Brida\\
        NIC, DESY, Platanenallee 6, 15738 Zeuthen, Germany\\
        E-mail: \email{mattia.dalla.brida@desy.de}}
        
\author{Mauro Papinutto\\
        Dipartimento di Fisica, "Sapienza" Universit\`a di Roma, and INFN, Sezione di Roma, Piazzale Aldo Moro 2, I-00185 Roma, Italy.\\
        E-mail: \email{mauro.papinutto@roma1.infn.it}}
        
\author{\speaker{ Pol Vilaseca}\\
        Instituto Nazionale di Fisica Nucleare (INFN), Sezione di Roma, P.le A. Moro 2, I-00185 Roma, Italy\\
        E-mail: \email{pol.vilaseca.mainar@roma1.infn.it}}

\abstract{The chirally rotated Schr\"odinger functional ($\chi$SF)  renders the mechanism of automatic $O(a)$ improvement compatible with Schr\"odinger functional (SF) renormalization schemes.
Here we define a family of renormalization schemes based on the $\chi$SF for a complete basis of $\Delta F = 2$ parity-odd four-fermion operators.
We compute the corresponding scale-dependent renormalization constants to one-loop order in perturbation theory and obtain their NLO anomalous dimensions by matching to the $\overline{\textrm{MS}}$ scheme.
Due to automatic $O(a)$ improvement, once the $\chi$SF is renormalized and improved at the boundaries, 
the step scaling functions (SSF) of these operators approach their continuum limit with $O(a^{2})$ corrections without the need of operator improvement.}

\FullConference{The 33rd International Symposium on Lattice Field Theory\\
		14 -18 July 2015\\
		Kobe International Conference Center, Kobe, Japan}

\begin{document}
\section{Introduction}
Flavour physiscs plays a fundamental role in the search for New Physics (NP) in particle accelerators. The existence
of new particles can be tested indirectly through the effects they might induce on processes at low-energies. Among 
flavour physics processes,
$\Delta F=2$ transitions provide very strong constrains on NP.  The most general $\Delta F=2$ weak effective Hamiltonian
can be constructed in terms of the following complete set of parity even (PE) and parity odd (PO) 4-quark operators,
\begin{align}
\textrm{PE}:&\qquad Q_{k}^{\pm} \in \left\{ Q_{VV+AA}^{\pm}, Q_{VV-AA}^{\pm},
Q_{SS-PP}^{\pm}, Q_{SS+PP}^{\pm},
2Q_{T\tilde T}^{\pm}\right\},\nonumber\\
\textrm{PO}:&\qquad \mathcal{Q}_{k}^{\pm} \in \left\{ \mathcal{Q}_{VA+AV}^{\pm},\mathcal{Q}_{VA-AV}^{\pm},
-\mathcal{Q}_{SP-PS}^{\pm}, \mathcal{Q}_{SP+PS}^{\pm},
2\mathcal{Q}_{T\tilde T}^{\pm}\right\}.
\label{eq:pe_po_ops}
\end{align}
Here, a 4-fermion operator with a particular Dirac structure and with four generic
flavours of quarks is given by
\begin{equation}
O_{XY}^{\pm} = {1\over2}\left[
(\overline\psi_{1}\Gamma_{X}\psi_{2})(\overline\psi_{3}\Gamma_{Y}\psi_{4})
\pm (2\leftrightarrow4)\right].
\end{equation}
From the operators in Eq.(\ref{eq:pe_po_ops}), only $Q_{1}$ and $\mathcal{Q}_{1}$ correspond to SM processes. All 
the others appear only in beyond SM processes.

Regularizations which break explicitly chiral symmetry generally induce complicated renormalization patterns 
for composite operators since they allow for
mixing among operators of different naive chirality. Indeed, when considering Wilson-fermions, 
all PE operators in Eq.(\ref{eq:pe_po_ops}) mix under renormalization. 
On the other hand, the PO operators renormalize
as in chirally preserving regularizations, namely \cite{M4}
\begin{equation}
[\mathcal{Q}_{1}]_{R}=\mathcal{Z}_{11}\mathcal{Q}_{1},\quad
  \left[ 
  {\begin{array}{c}
   \mathcal{Q}_{2}  \\   
   \mathcal{Q}_{3} \      
   \end{array} }
    \right]_{R}=
      \left[ 
  {\begin{array}{cc}
   \mathcal{Z}_{22} & \mathcal{Z}_{23} \\   
   \mathcal{Z}_{32} & \mathcal{Z}_{33}\      
   \end{array} }
    \right]
      \left[
  {\begin{array}{c}
   \mathcal{Q}_{2}  \\   
   \mathcal{Q}_{3} \      
   \end{array} }
    \right],\quad
  \left[
  {\begin{array}{c}
   \mathcal{Q}_{4}  \\   
   \mathcal{Q}_{5} \      
   \end{array} }
    \right]_{R}=
      \left[ 
  {\begin{array}{cc}
   \mathcal{Z}_{44} & \mathcal{Z}_{45} \\   
   \mathcal{Z}_{54} & \mathcal{Z}_{55}\      
   \end{array} }
    \right]
      \left[ 
  {\begin{array}{c}
   \mathcal{Q}_{4}  \\   
   \mathcal{Q}_{5} \      
   \end{array} }
    \right].
    \label{eq:po_ren}
\end{equation}
In practice, one can avoid the renormalization of PE operators following the strategies in \cite{M5} and \cite{M6}, for which 
only renormalized matrix elements of PO operators are needed.

In this work we thus focus on the renormalization of PO operators. 
We first construct a suitable set of 3-point functions in
the $\chi$SF in order to define the renormalization conditions. Thanks to the mechanism of automatic $O(a)$ improvement all renormalization factors considered will be affected only by $O(a^{2})$ lattice artefacts,
even without $O(a)$ improving the operators or the action in the bulk. We then expand the renormalization conditions 
obtained to 1-loop order in perturbation theory, and perform an exploratory study that sets the basis for future non-perturbative studies. 

\section{A few words on the $\chi$SF}
Chirally rotated boundary conditions for a flavour doublet of fermionic fields $\psi$ and $\overline{\psi}$ take the form \cite{XSF}
\begin{equation}
 \left.\widetilde{Q}_{+}\psi(x)\right|_{x_{0}=0} =  \left.\widetilde{Q}_{-}\psi(x)\right|_{x_{0}=T} =0,
\qquad \left.\overline{\psi}(x)\widetilde{Q}_{+}\right|_{x_{0}=0} = \left.\overline{\psi}(x)\widetilde{Q}_{-}\right|_{x_{0}=T}=0,
\label{eq:boundary_conditions}
\end{equation}
with the projectors $\widetilde{Q}_{\pm}=\frac{1}{2}(1\pm i\gamma_{0}\gamma_{5}\tau^{3})$ and where $\tau^{i}$ are
the Pauli matrices. 
These boundary conditions are related to the standard SF boundary
conditions through the non-anomalous chiral rotation
\begin{equation}
 \psi\rightarrow R(\alpha)\psi,\quad \overline{\psi}\rightarrow\overline{\psi}R(\alpha),
\quad R(\alpha)=\exp(i\alpha\gamma_{5}\tau^{3}/2),
\quad \alpha = \pi/2.
\label{eq:rotation}
\end{equation}
A dictionary relating correlation functions in both setups can be established through the mapping
\begin{equation}
 \langle\mathcal{O}[\psi,\overline{\psi}]\rangle_{\chi\textrm{\small SF}}=
 \langle\mathcal{O}[R(-\pi/2)\psi,\overline{\psi}R(-\pi/2)]\rangle_{\textrm{\small SF}}.
\label{eq:corr_rel}
\end{equation}

The boundary conditions in Eq.(\ref{eq:boundary_conditions}) are invariant under the rotated version of 
parity\footnote{$P_{5}:\psi(x)\rightarrow i\gamma_{0}\gamma_{5}\tau^{3}\psi(\widetilde{x}),\quad 
P_{5}:\overline{\psi}(x)\rightarrow -\overline{\psi}(\widetilde{x})i\gamma_{0}\gamma_{5}\tau^{3},\quad
\widetilde{x}=(x_{0},-{\bf x})$} 
$P_{5}$,
i.e. $[\widetilde{Q}_{\pm},\gamma_{0}\gamma_{5}\tau^{3}]=0$.
The $P_{5}$ transformation can thus be used to distinguish between 
$P_{5}$-even and $P_{5}$-odd correlation functions and invoke the mechanism
of automatic $O(a)$ improvement: on the lattice, all bulk $O(a)$ effects are  absent from $P_{5}$-even correlation functions, while these are contained in the $P_{5}$-odd observables which are thus pure lattice artefacts.
For automatic $O(a)$ improvement to be at work one needs to set the quark masses to their critical value and tune the
coefficient of a dimension 3 boundary counterterm, in order to obtain the correct symmetries of
the $\chi$SF in the continuum limit.
Note that additional $O(a)$ effects originating from the boundaries are present in any correlation function. These can be eliminated introducing
a couple of $O(a)$ counterterms at the space-time boundaries.  
%
In the following, we consider the lattice set-up described in \cite{PV2014,DB2014}, to where we refer for details on 
the action and on the renormalization and improvement conditions for determining the critical mass and boundary counterterms.

\section{Correlation functions of 4-fermion operators in the $\chi$SF}
In the following we define a set of correlation functions for PO 4-fermion operators in the $\chi$SF. These are 
obtained by rotating correlation functions of PE operators in the standard SF through Eq.(\ref{eq:corr_rel}).
We start defining two different types of 3-point correlation functions  in the standard SF, with an arbitrary
choice of flavours,
\begin{equation}
F_{i}(x_{0})=\langle O_{5}^{'21}Q_{i}^{1234}(x_{0})O_{5}^{43} \rangle,\qquad
K_{i}(x_{0})={1\over3}\sum_{k=1}^{3}\langle O_{k}^{'21}Q_{i}^{1234}(x_{0})O_{k}^{43} \rangle,
\label{eq:SF_4f_corrs}
\end{equation}
where $i\in[1,5]$ labels the basis of PE operators in Eq.(\ref{eq:pe_po_ops}) and where
$O_{5}$, $O'_{5}$, $O_{k}$ and $O'_{k}$ are standard SF boundary operators.
In the above equations we choose the flavour combination $f_{1}=f_{2}=f_{4}=u$ and $f_{3}=d$. 
After applying the chiral rotation Eq.(\ref{eq:rotation}) one obtains the mapping
\begin{align}
Q_{1}^{\pm}\rightarrow -i \mathcal{Q}_{1}^{\pm,uudu},\textrm{ }
Q_{2}^{\pm}\rightarrow -i \mathcal{Q}_{2}^{\mp,uudu},\textrm{ }
Q_{3}^{\pm}\rightarrow -i \mathcal{Q}_{3}^{\mp,uudu},\textrm{ }
Q_{4}^{\pm}\rightarrow -i \mathcal{Q}_{4}^{\pm,uudu},\textrm{ }
Q_{5}^{\pm}\rightarrow -i \mathcal{Q}_{5}^{\pm,uudu}.
\end{align}
Hence, by applying the rotation Eq.(\ref{eq:rotation}) to Eq.(\ref{eq:SF_4f_corrs}) we obtain correlation functions of 
PO operators in the $\chi$SF,
\begin{align}
F_{i}(x_{0})\rightarrow G_{i}(x_{0})=\langle \mathcal{O}_{5}^{'uu}\mathcal{Q}_{i}^{uudu}(x_{0})\mathcal{O}_{5}^{ud} \rangle,\quad
K_{i}(x_{0})\rightarrow L_{i}(x_{0})={1\over3}\sum_{k=1}^{3}\langle \mathcal{O}_{k}^{'uu}\mathcal{Q}_{i}^{uudu}(x_{0})\mathcal{O}_{k}^{ud} \rangle.
\label{eq:XSF_4f_corrs}
\end{align}
Here $\mathcal{O}_{5}^{12}$, $\mathcal{O}_{5}^{'12}$, $\mathcal{O}_{k}^{12}$ and $\mathcal{O}_{k}^{'12}$
are the boundary fermion bilinears in the $\chi$SF \cite{DB2016}. 
We also consider the boundary to boundary correlation functions $g_{1}^{12}$ and $l_{1}^{12}$, and the 
boundary to bulk correlation functions $g_{\tilde{V}}^{12}(x_{0})$ and $l_{\tilde{V}}^{12}(x_{0})$,
where $\tilde{V}_{\mu}^{12}$ is the conserved vector current.  These are renormalized\footnote{
Note that since the vector current $\tilde{V}_{\mu}^{12}$ is exactly conserved, there is no renormalisation factor for the fermion bilinear in $[g_{\tilde V}^{12}]_{R}$ and $[l_{\tilde V}^{12}]_{R}$.} by multiplying them
with the proper power of renormalization factors of the boundary fields $Z_{\zeta}$, i.e.
\begin{align}
[g_{1}^{12}]_{R}=Z_{\zeta}^{4}g_{1}^{12},\quad
 [l_{\tilde V}^{12}]_{R}=Z_{\zeta}^{2}l_{\tilde V}^{12},\quad
  [g_{\tilde V}^{12}]_{R}=Z_{\zeta}^{2}g_{\tilde V}^{12},\quad
[l_{1}^{12}]_{R}=Z_{\zeta}^{4}l_{1}^{12}.
\label{eq:ren_g}
\end{align}

Due to the presence of the boundary fields in Eq.(\ref{eq:XSF_4f_corrs}), the renormalized $G_{i}$ and $L_{i}$ also need the renormalization factor $Z_{\zeta}$. This can be eliminated by normalising $G_{i}$ and $L_{i}$ with a suitable combination of boundary to boundary or boundary to bulk correlation functions. To this end, we define the ratios
\begin{equation}
\mathcal{G}_{i} = G_{i}/\mathcal{N}(\alpha,\beta,\gamma),\qquad
\mathcal{L}_{i} = L_{i}/\mathcal{N}(\alpha,\beta,\gamma).
\label{eq:G_L_norm}
\end{equation}
The normalization $\mathcal{N}(\alpha,\beta,\gamma)$ is given by
\begin{equation}
\mathcal{N}(\alpha,\beta,\gamma) =
\left\{ {g_{1}^{ud}\over g_{1}^{ud(0)}}\right\}^{\alpha}
\left\{ {l_{1}^{ud}\over l_{1}^{ud(0)}}\right\}^{\beta}
\left\{ {g_{\tilde V}^{ud}\over g_{\tilde V}^{ud(0)}}\right\}^{2\gamma}
\left\{ {l_{\tilde V}^{uu}\over l_{\tilde V}^{uu(0)}}\right\}^{2(1-\alpha-\beta-\gamma)},
\label{eq:normalization}
\end{equation}
where $\alpha$, $\beta$ and $\gamma$ are real coefficients satisfying the condition $\alpha+\beta+\gamma \le 1$,
and where the superscript $(0)$ denotes the tree-level correlation functions.  
The exponents in the different
terms of Eq.(\ref{eq:normalization}) are chosen such that the correct power of the renormalization
factor $Z_{\zeta}$ is eliminated in Eq.(\ref{eq:G_L_norm}).

\section{Renormalization conditions}
Renormalization conditions can be obtained by demanding some renormalized correlation functions to be equal
to their tree level values at a given scale $\mu=L^{-1}$, where $L$ is the size of the system in the spatial directions. In order to specify a particular renormalization scheme
one has to: a) choose between the observables $\mathcal{G}_{i}^{\pm}$
and $\mathcal{L}_{i}^{\pm}$ for imposing the renormalization conditions, b) fix the parameters 
$\alpha$, $\beta$ and $\gamma$, and c) fix the dimensionless parameters $\theta$, $\rho=T/L$
and the timeslice $x_{0}$ at which correlations are evaluated. 
In the following we will always consider
$\rho=1$ and $x_{0}=T/2$. 

It is convenient to write explicitly the parameter dependance on the matrix elements in Eq.(\ref{eq:G_L_norm})
and introduce the following notation,
\begin{equation}
\mathcal{H}_{i,{\bf c}}^{1,\pm}(\theta) = \mathcal{G}_{i}^{\pm}(x_{0}=T/2),\qquad
\mathcal{H}_{i,{\bf c}}^{2,\pm}(\theta) = \mathcal{L}_{i}^{\pm}(x_{0}=T/2),\qquad
{\bf c}=(\alpha,\beta,\gamma).
\label{eq:H1_H2}
\end{equation}
This way, renormalization conditions for the operator which does not mix are specified by
\begin{equation}
\mathcal{Z}_{11}^{\pm}(g_{0},a\mu)\mathcal{H}_{1,{\bf c}}^{s,\pm}(\theta) 
= \left.\mathcal{H}_{1,{\bf c}}^{s,\pm}(\theta) \right|_{g_{0}=0},
\qquad s=1,2.
\label{eq:Q1_cond}
\end{equation}
For the pairs of operators which mix, we impose renormalization conditions on each 
$2\times2$ matrix by forming combinations of the form \cite{MP2014}
\begin{equation}
  \left( 
  {\begin{array}{cc}
   \mathcal{Z}_{22} & \mathcal{Z}_{23} \\   
   \mathcal{Z}_{32} & \mathcal{Z}_{33}\      
   \end{array} }
    \right)
      \left( 
  {\begin{array}{cc}
   \mathcal{H}_{2,{\bf c}_{1}}^{s_{1}}(\theta_{1}) & \mathcal{H}_{2,{\bf c}_{2}}^{s_{2}}(\theta_{2}) \\   
   \mathcal{H}_{3,{\bf c}_{3}}^{s_{1}}(\theta_{1}) & \mathcal{H}_{3,{\bf c}_{4}}^{s_{2}}(\theta_{2})\      
   \end{array} }
    \right)=
    \left.\left( 
  {\begin{array}{cc}
   \mathcal{H}_{2,{\bf c}_{1}}^{s_{1}}(\theta_{1}) & \mathcal{H}_{2,{\bf c}_{2}}^{s_{2}}(\theta_{2}) \\   
   \mathcal{H}_{3,{\bf c}_{3}}^{s_{1}}(\theta_{1}) & \mathcal{H}_{3,{\bf c}_{4}}^{s_{2}}(\theta_{2})\      
   \end{array} }
    \right)\right|_{g_{0}^{2}=0},
    \label{eq:mat_cond}
\end{equation}
and similarly with the conditions for the operators $\mathcal{Q}_{4}$ and $\mathcal{Q}_{5}$. 
In order to define sensible renormalization conditions the tree-level matrix of the r.h.s of
Eq.(\ref{eq:mat_cond}) must be invertible. Note that the source type, the parametes $\bf{c}$ and $\theta$ must be the same within the different columns.

The amount of parameters appearing in the renormalization conditions Eqs.(\ref{eq:Q1_cond})
and (\ref{eq:mat_cond}) leaves quite some freedom in choosing particular renormalization schemes.
It has been pointed out \cite{MP2014} that the anomalous dimensions (AD) of the four fermion operators that mix 
can be large.
A possible criteria to choose a renormalization scheme is to look for those for which the ADs are not too large. 
At the present order In perturbation theory this is equivalent to look for combinations of parameters 
that lead to an RG-evolution for which the NLO contribution is close to the LO.
With this in mind, we describe the 
RG-evolution of a matrix $\mathcal{Z}$ of Z-factors
$\mathcal{Z}(\mu_{1})=U(\mu_{1},\mu_{2})\mathcal{Z}(\mu_{2})$ as \cite{MP2014}
\begin{equation}
U(\mu_{2},\mu_{1})=\textrm{T}\exp\left\{
\int_{\overline{g}(\mu_{1})}^{\overline{g}(\mu_{2})}
\frac{\gamma(g)}{\beta(g)}dg
\right\}\equiv\tilde{U}^{-1}(\mu_{1})\tilde{U}(\mu_{2}).
\label{eq:formal_RG}
\end{equation}
The matrix $\tilde{U}(\mu)$ is expanded in perturbation theory as
\begin{equation}
\tilde{U}(\mu)=
\left[\frac{\overline{g}^{2}(\mu)}{4\pi}\right]^{-\frac{\gamma_{0}}{2\beta_{0}}}
\left[\textbf{1}+\overline{g}^{2}(\mu)J(\mu)+O(\overline{g}^{4})\right],
\end{equation}
where the matrix $J(\mu)$ contains the NLO evolution and satisfies
\begin{equation}
\frac{\partial}{\partial\mu}J(\mu)=0,\qquad
J-\left[\frac{\gamma_{0}}{2\beta_{0}},J\right]=
\frac{\beta_{1}}{2\beta_{0}^{2}}\gamma_{0}-\frac{1}{2\beta_{0}}\gamma_{1}.
\label{eq:J_eqs}
\end{equation}
The matrix $J(\mu)$ depends explicitly on the AD at NLO $\gamma_{1}$. A criteria in choosing a particular
renormalization scheme is thus to demand the norm of $J$ to be as small as possible. 
This is equivalent to ask the NLO scale evolution of the Z-factors to be as close as possible to the LO evolution. 
One can thus choose appropriate combinations
of observables and parameters in Eqs.(\ref{eq:Q1_cond}) and (\ref{eq:mat_cond}) to minimize the norm of $J$.
Schemes obtained following this strategy are desirable in the matching at high energy.
On the other hand, minimizing $|J|$ does not guarantee a good behaviour of the perturbative expansion. 
Ultimately, only the comparison of the non-perturbative and perturbative
evolutions will determine which schemes allow for a reliable matching at high-energy. 
The idea is thus to use 1-loop perturbation theory to identify a set of potentially 
good schemes and most importantly avoid schemes with particularly large NLO
ADs. Finally, having several different candidate schemes allows to compare the RGI
operators obtained and thus have an estimate on the goodness of the matching.


\section{Results}
A given renormalization factor $\mathcal{Z}_{ij}$ in Eq.(\ref{eq:po_ren}) is expanded in perturbation theory as
$\mathcal{Z}_{ij}\simeq 1 + \overline{g}^{2}\mathcal{Z}_{ij}^{(1)}+O(\overline{g}^{4})$. The 1-loop coefficient $\mathcal{Z}^{(1)}_{ij}$
has an asymptotic form in $a/L$ given by
\begin{equation}
\mathcal{Z}^{(1)}_{ij}= \sum_{n=0}^{\infty}\left(r_{n,ij}+s_{n,ij}\ln(a/L)\right)(a/L)^{n}.
\label{eq:Z_as}
\end{equation}
Here, $r_{0,ij}$ is the finite asymptotic part of $\mathcal{Z}^{(1)}_{ij}$, $s_{0,ij}$ is the universal LO anomalous dimension,
and the $r_{n,ij}$ and $s_{n,ij}$ coefficients with $n>0$ correspond to $O(a^{n})$ cutoff effects. Due to automatic $O(a)$ 
improvement, the coefficients $s_{1,ij}$ should be zero regardless of whether the action or the operators have been improved in the bulk. The coefficients $r_{1,ij}$ are zero once boundary $O(a)$ improvement is implemented. 

\begin{figure}[!ht]
\centering
\includegraphics[clip=true,scale=0.5]{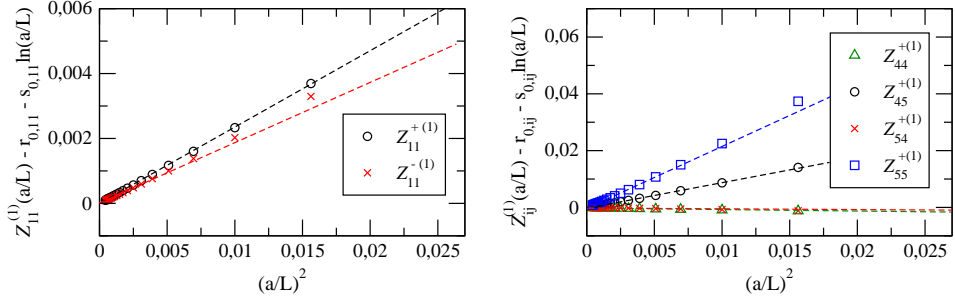}
\caption{Convergence to the continuum limit of the 1-loop renormalization factors $\mathcal{Z}^{\pm,(1)}_{11}$ 
(left pannel) and $\mathcal{Z}^{+,(1)}_{44}$, $\mathcal{Z}^{+,(1)}_{45}$, $\mathcal{Z}^{+,(1)}_{54}$ 
$\mathcal{Z}^{+,(1)}_{55}$ (right pannel), for a specific scheme choosen for illustrative purposes, after substracting the corresponding finite parts and logarithmic divergencies. The 
coloured discontinuous lines are fits to the data excluding the coarsest lattice spacings. }
\label{fig:dZ}
\end{figure}

In order to calculate the different $\mathcal{Z}^{(1)}_{ij}$ we expand 
Eqs.(\ref{eq:Q1_cond}) and (\ref{eq:mat_cond}) to 
$O(\overline{g}^{2})$ in perturbation theory 
and compute numerically the Feynman diagrams contributing to the different correlation functions involved. 
Details on the gauge fixing procedure and on the expansion of the correlation functions will be 
reported elsewhere. In this way we obtain explicit expressions for the coefficients $\mathcal{Z}^{(1)}_{ij}$ in terms of the
parameters $\bf{c}$ \footnote{The rest of the parameters appearing in Eqs.(\ref{eq:Q1_cond}) and (\ref{eq:mat_cond}) must be fixed explicitely.}.
From this, it is easy to apply the criteria introduced in the previous section to find appropriate renormalization schemes. For all operators we 
obtain the correct value of the LO anomalous dimension, which is a strong check on the calculation. Moreover, we find all
$r_{1,ij}$ and $s_{1,ij}$ coefficients to be consistent with 0, confirming the absence of $O(a)$ effects.
Examples for the determination of the finite part for some schemes considered are shown in Figure  \ref{fig:dZ}.

For the opeators $\mathcal{Q}_{1}^{\pm}$, Eqs.(\ref{eq:formal_RG} - \ref{eq:J_eqs}) are scalar and it is
possible to find schemes for which $J(\mu)$ is as small as desirable. 
For each pair of operators that mix, we build renormalization conditions considering all possible combinations
of observables with either $\theta=0.0$ or $0.5$. For each condition obtained in this way, we find the sets 
of parameters $(\alpha_{i},\beta_{i},\gamma_{i})$ that minimize  $|J|$. Since the value of $|J|$ remains stable in the 
neighbourhood of the minimum, we can identify regions in parameter space for which the NLO
contributions are not unreasonably large. Any choice of parameters outside these regions 
can otherwise lead to arbitrarily large NLO contributions.

\begin{figure}[!ht]
\centering
\includegraphics[clip=true,scale=0.5]{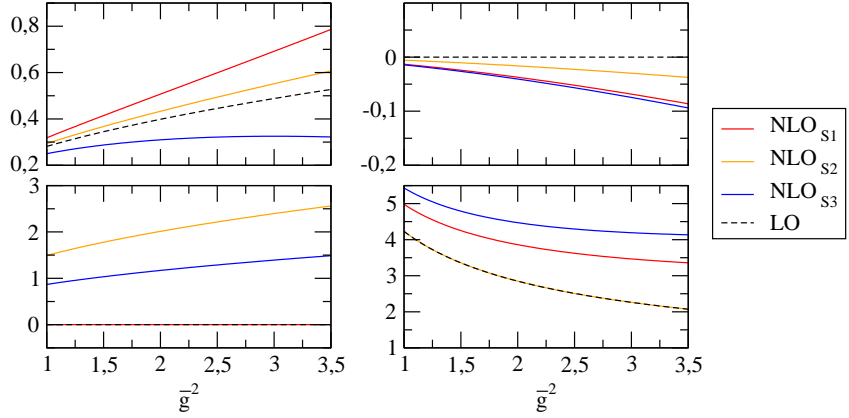}
\caption{Components of tha matrix $\tilde{U}(\mu)$ in the basis in which $\gamma_{0}$ is diagonal. as a function of $\overline{g}^{2}$ for the operators $\{\mathcal{Q}^{+}_{4},\mathcal{Q}^{+}_{5}\}$, in three different renormalization schemes.}
\label{fig:U_running}
\end{figure}

As an example, 
in Figure \ref{fig:U_running} we show the LO and NLO
evolution of $\tilde{U}$ in
terms of $\overline{g}^{2}$, in the basis in which $\gamma_{0}$ is diagonal, for 3 schemes $S_{i}$ ($i=1,2,3$)
for the pair $\left\{\mathcal{Q}_{4}^{+},\mathcal{Q}_{5}^{+}\right\}$. The values of $(\alpha_{i},\beta_{i},\gamma_{i})$
have been chosen to minimize $|J|$ in the three schemes. The LO is universal, with only the diagonal
terms being non-zero in the chosen basis. 
The difference between LO and NLO evolution is more evident in some of the matrix elements, but which
matrix element differs the most depends on the scheme. The overall behaviour of the scale evolution looks however
comparable for all schemes once
$(\alpha_{i},\beta_{i},\gamma_{i})$ are chosen close to the values which minimise the norm of $J$.

\section{Conclusions}
In this work we have set up the $\chi$SF for the renormalization of parity odd $\Delta F=2$ 4-fermion operators.
We have defined a set of 3 point correlation functions for building renormalization conditions. From these, we expect 
an improvement of the statistical error
 in non-perturbative calculations with respect to the standard SF, where 4 point functions
have to be used \cite{MP2014}.
Thanks to automatic $O(a)$ improvement, bulk operator improvement can be avoided. Likely, this will allow to extrapolate the lattice SSF of these operators as $O(a^{2})$, increasing significantly the precision on the continuum results.
The flexibility of our renormalization conditions allows us to seek for schemes for which the NLO is as close as possible to the LO RG-evolution. In fact, we were able to identify potentially good schemes in this respect. The strategy developed seems promising, and deeper studies in parameter space are on the way. Ultimately of course only a full non-perturbative study can show which are the most suitable schemes for the determination of the relevant RGI operators. An additional important point to investigate is the size of cutoff effects in the corresponding SSFs. These results will then set the ground for future non-perturbative studies using the $\chi$SF.

Numerical calculations were performed at the Galileo machine at CINECA. We gratefully thank Stefan Sint and Tassos Vladikas for insightful discussions.


\begin{thebibliography}{99}


\bibitem{M4}
  A.~Donini {\it et al.}
  Eur.\ Phys.\ J.\ C\  {\bf 10} (1999) 121.
\bibitem{M5}
  R.~Frezzotti {\it et al.} 
  JHEP  {\bf 0108} (2001) 058;
  R.~Frezzotti and G.~C.~Rossi, 
  JHEP {\bf 0410} (2004) 070.
 \bibitem{M6}
 D.~Becirevi\'c {\it et al.} 
 Phys. Lett. B {\bf 487} (2000) 74.
 \bibitem{DB2016}
  M.~Dalla Brida,~S.~Sint,~P.~Vilaseca,
 arXiv:1603.00046 (2016).
 \bibitem{XSF}
  S.~Sint,
  Nucl.\ Phys.\ B.\  {\bf 847} (2011) 491,
\bibitem{PV2014}
  S.~Sint,~P.~Vilaseca,
  PoS LATTICE {\bf2014}, (2014) 279.
\bibitem{DB2014}
  M.~Dalla Brida,~S.~Sint,
  PoS LATTICE {\bf2014}, (2014) 280.
  \bibitem{MP2014}
  M.~Papinutto,~C.~Pena,~D.~Preti,
  PoS LATTICE {\bf2014}, (2014) 281.




\end{thebibliography}
\end{document}